# Proposed 2MW Wind Turbine for Use in the Governorate of Dhofar at the Sultanate of Oman

Osama Ahmed Marzouk, Omar Rashid Hamdan Al Badi, Maadh Hamed Salman Al Rashdi, Hamed Mohammed Eid Al Balushi

College of Engineering, University of Buraimi, Al Buraimi, Sultanate of Oman

**Email address:**
osama.m@uob.edu.om (O. A. Marzouk)

**Abstract:** In this work, we propose a preliminary design of a horizontal-axis wind turbine (HAWT) as a candidate for the Dhofar Wind Farm project, in the southern Omani Governorate "Dhofar", at the southwest part of the Sultanate of Oman. This wind farm (under construction) is considered to be the first commercial, utility-scale (50MW) wind farm in the GCC (Gulf Cooperation Council) area. The proposed wind turbine has an expected electricity generation of 2MW. We studied the wind atlas of Oman and from which we determined the maximum possible mean wind speed in the entire Sultanate and built our design based on that reference value, which is 6m/s (21.6km/h). After this, we applied a set of modeling equations that estimate the power output from the wind turbine rotor and matched the target electric power to the design variables using a MATLAB computer code. We reached a suitable design and we present here the distribution of the blade angle (twist angle), and the power per unit span along the rotor blade. The rotor design has 3 blades with a diameter of 70m and a rotational speed of 24rpm. This rotor gives 2.37MW of output power, which exceeds the target 2MW output, allowing for about 15% of power losses in the gearbox and generator. We utilized some commercial designs of wind turbines from different international manufacturers as references for typical limits or recommended values of some design parameters.

## 1. Introduction

For thousands of years, humans have been using wind energy. Examples that still exist today include sailing boats and windmills to grind grains. In 1700B. C., King Hammurabi of Babylon used wind power for irrigation [1, 2], where winds hit rotating bowls that pump water. Winds were used for the first time to generate electricity in 1888 in the USA by Charles Brush who invented the wind turbine [3]. The diameter of its rotor was 17m and it had wooden blades. Its power output was only 12kW. Nowadays, typical wind turbines are much larger and more powerful because of better materials, advances in aerodynamics as well as years of experience.

As shown in Figure 1, the world need for energy (electricity) is increasing [4] as a result of growth in the residential, commercial, industrial, and transportation sectors. Although wind and solar powers are not a big source of energy currently, they show the largest growth over years compared to other energy sources.

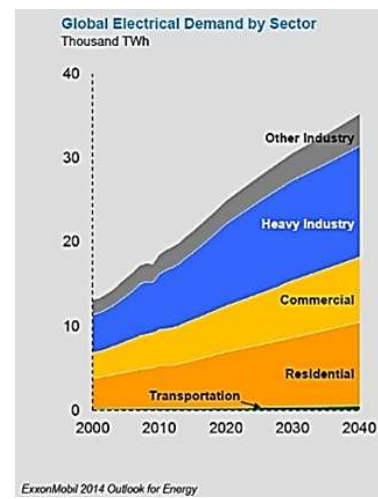

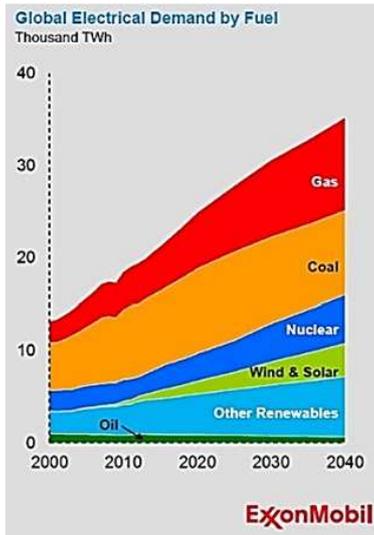

*Figure 1. Predicted energy demands and generations in the world until 2040.*

Figure 2 shows a historical record of the world energy consumption (by fuel) over about two centuries. One can see that the consumption is overall growing with an increasing rate.

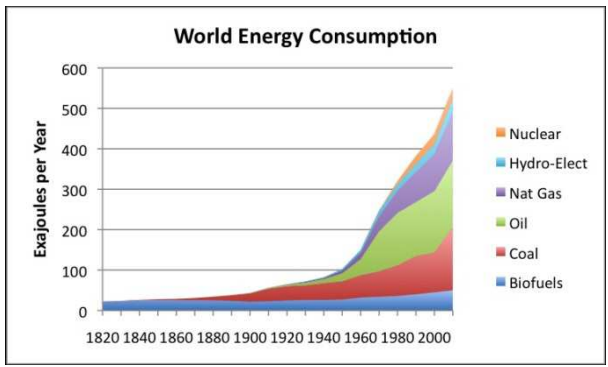

*Figure 2. World energy consumption, based on Vaclav Smil estimates from Energy Transitions: History, Requirements and Prospects together with BP Statistical Data for 1965 and subsequent. Published at: http://theoildrum.com/node/9023*

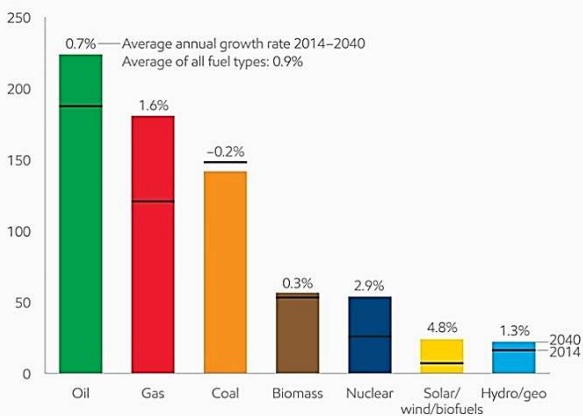

*Figure 3. Global fuel demand in 2014 (actual) and 2040 (predicted).*

Figure 3 [5] compares the actual global fuel demand in 2014 and the expected one in 2040. Renewable sources (including wind energy) have the highest expected growth rate (2014-2040) of 4.8%.

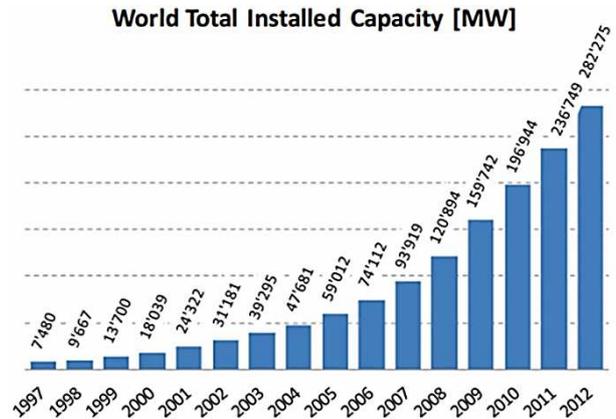

*Figure 4. World total installed capacity of wind energy.*

In particular the demand on clean renewable energy is increasing given different factors such as concerns about pollution from fossil fuels and also the limited (although huge) reserves of these fossil fuels. Figure 4 [6] shows an accelerating utilization of wind energy in terms of installed capacity. Over 10 years only (from 2003 to 2012), it increased from 39.3GW to 282.3GW, i.e., by a factor of about 7.2. Over the last 30 years, wind turbines have shown very fast increase in size and efficiency, with rates comparable to those observed in computer technology advancement [1].

According to "Vision 2020" of His Majesty Sultan Qaboos bin Said (ruler of the Sultanate of Oman), Oman is moving towards providing 10% of its electricity needs from renewal energy sources by 2020. Therefore, Oman plans to invest a lot in wind energy. Oman is the only dedicated country in the region having a Ministry for Environment and Climate Affairs [7].

The ongoing interest in wind energy in Oman is proven by the planning of the first wind farm in the Sultanate and the entire GCC area, with an estimated capacity of 50MW. Both the Rural Areas Electricity Company (Tanweer, formerly "RAECO") in Oman and Abu Dhabi Future Energy Company (Masdar) in the United Arab Emirates (UAE), supported this project, which is fully funded by the Abu Dhabi Fund for Development (ADFD) [8]. The project was planned to be constructed in Thumrait, located in the Dhofar Governorate in the southwest part of Oman. This Dhofar Wind Farm project shall demonstrate that wind energy is achievable in Oman. The number of turbines to be installed in an initial view was to vary between 17 and 25 according to the generation capacity of each turbine, such that the total production is 50MW. However, the final decision was to have 13 wind turbines, at 3.8MW each. The first turbine was installed near the end of December 2018 [9]. One of the 13 wind turbines was commissioned and the farm was successfully connected to Oman's electricity transmission grid during the first half of August 2019 [8]. The project received consulting and engineering services by an Austrian-based international firm: ILF Consulting Engineers [10]. The expected generated electricity is enough for about 16,000 homes [11]. This represents 7% of the total power demand for Dhofar Governorate, with an expected offset of about 110,000 tonnes of carbon dioxide emissions per year. The project was initially planned to be in operation in 2017 but delays occurred. It is hoped that the whole farm becomes commercially operational by the end of 2019 [12].

## 2. Oman Wind Data

We can benefit a lot from the existing work by others in the same area. An important component we needed is the distribution of the mean wind data in Oman, (or wind atlas). A team at Sultan Qaboos University in Muscat conducted a useful research about the wind speed in the Sultanate of Oman. This research work [13] includes annual mean wind speeds in several Omani areas and wind directions. The data there are based on hourly wind data collected over five years, from 2004 to 2008 at 29 meteorological stations. These data were provided by the Directorate General of Meteorology and Air Navigation (DGMAN) in Oman. The stations are listed in Table 1, arranged in an alphabetical order.

***Table 1.*** *Names of meteorological stations analyzed by Reference [13].*

| 1 | Adam | 11 | Khasab | 21 | Rustaq |
|---|---|---|---|---|---|
| 2 | Bahla | 12 | Marmul | 22 | Saiq |
| 3 | Buraimi | 13 | Masirah | 23 | Salalah |
| 4 | Diba | 14 | Mina Qaboos | 24 | Samail |
| 5 | Duqm | 15 | Mina Salalah | 25 | Seeb |
| 6 | Fahud | 16 | Nizwa | 26 | Sohar |
| 7 | Ibra | 17 | Qalhat | 27 | Sur |
| 8 | Ibri | 18 | Qaran Alam | 28 | Thumrait |
| 9 | Jabal Shams | 19 | Qayroon Hyriti | 29 | Yalooni |
| 10 | Joba | 20 | Ras Alhad | | |

A map of Oman showing these stations is given in Figure 5, which is taken from the published work itself. The map is colored by the elevation.

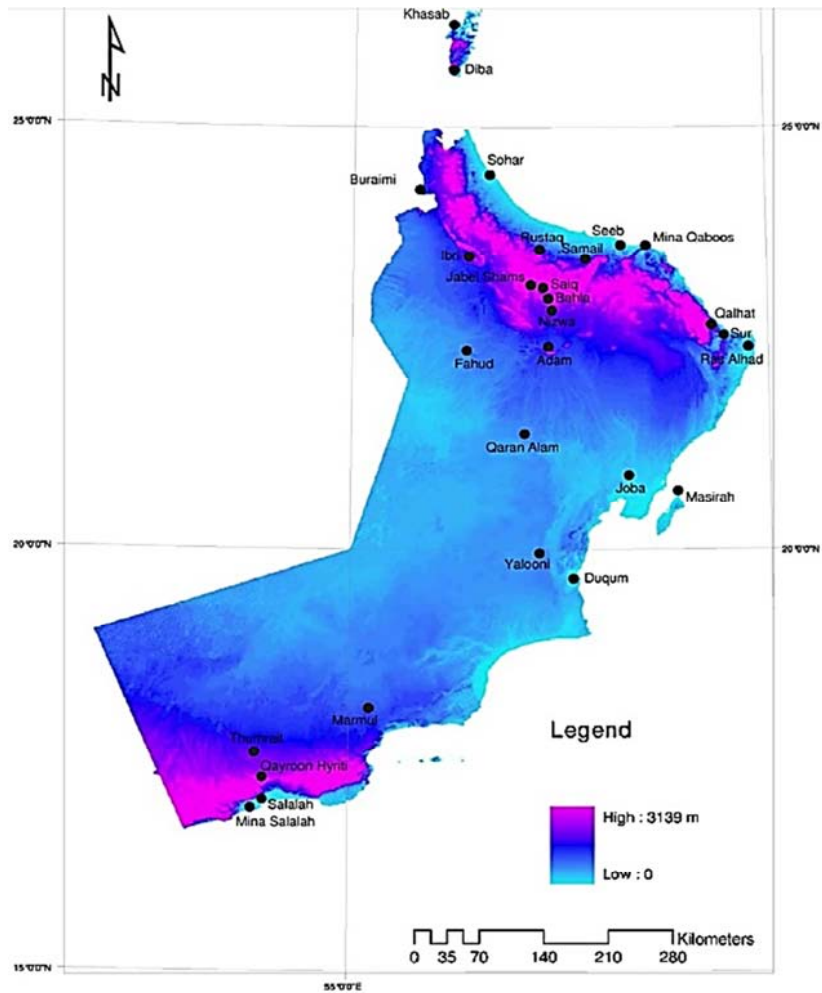

***Figure 5.*** *Location of the meteorological stations analyzed by Reference [13].*

Figure 6 [13] shows the annual and seasonal mean wind speeds at the 29 different locations in Oman. From the figure, it becomes clear that Thumrait is one of the best places for installing a wind turbine because the wind speed is highest, being about 6m/s (21.6km/h).

Although Qayroon Hyriti has a similar mean annual wind speed, Thumrait is preferred because it is located at a lower elevation, which means a larger air density and thus a higher wind power available. However, one should note that the wind in Thumrait shows more variation over the year than Qayroon Hyriti, where the wind is strongest in summer and reaches 7.5m/s. On the other hand, the wind there becomes weakest the winter and drops to 4m/s. So, the seasonal change in the wind speed in Thumrait is 3.5m/s, which is 58% of the recorded annual mean of 6m/s.

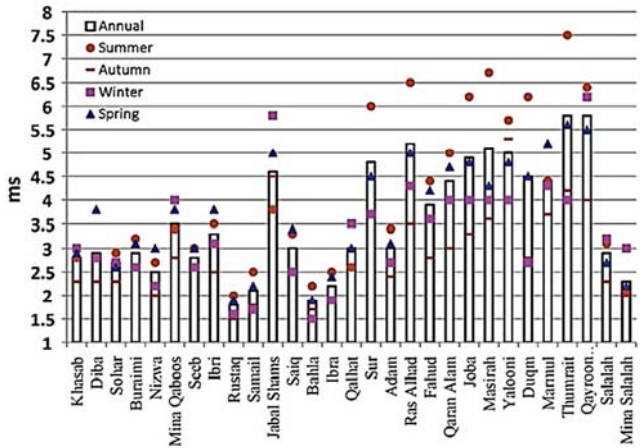

*Figure 6. Annual and seasonal mean wind speed at 29 different locations in Oman.*

This maximum observed wind speed in Oman of 6m/s is considered moderately suitable for a wind energy system [14]. We shall use this value as a fixed input in our design calculations.

Figure 7 is the wind rose of Thumrait in the summer (where the wind is strongest). It shows that the wind comes mostly from the South and there is little direction change. This is a good point and again makes Thumrait a preferred location.

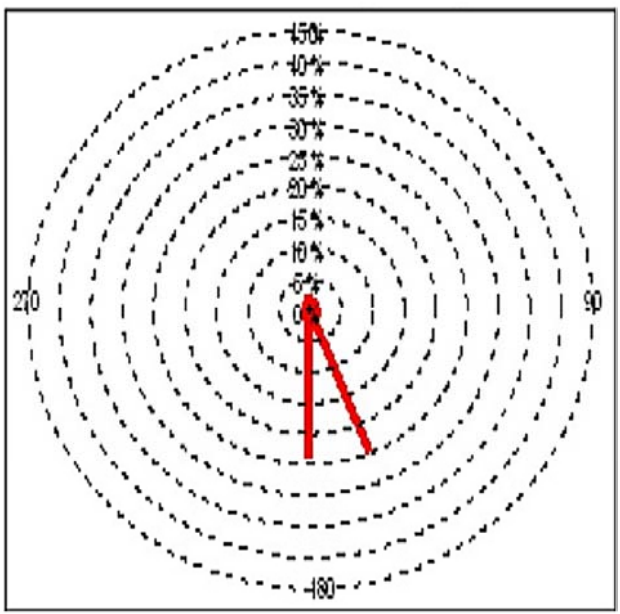

*Figure 7. Wind rose in Thumrait for the summer season.*

## 3. Benchmarking Designs

To obtain guidelines for design values for a typical wind turbine of the target capacity of 2MW, we utilized published online data from 4 different manufactures of wind turbines and the details of some of their wind turbine models of that capacity were examined. These manufactures companies are:
1. DeWind (Germany and South Korea).
2. ENERCON (Germany).
3. General Electric, GE (USA).
4. Vestas (Denmark).

## 4. Mathematical Modeling

This section covers the equations used to estimate the power output from a wind turbine [15]. The power output here is the power delivered by the rotor. Therefore, this calculated power is higher than the electricity generated from the wind turbine because it does not include losses in components like the gearbox and the generator. The equations given here are to be applied in the next section to give numerical results for the specific design we suggest for the wind turbine, but here they are only given in a suitable sequence to show the calculation steps.

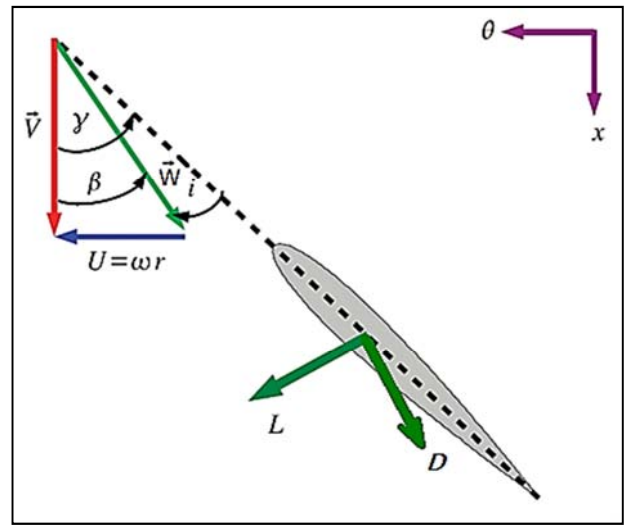

*Figure 8. Explanation of flow angles and velocities.*

Figure 8 is important to understand before presenting the equations. It explains the velocity triangle of the wind (velocity values and angles), and the resulting lift force on the airfoil of the rotor blade because of the wind effect and aerodynamics of its flow. The reader is advised to consult the Nomenclature of this work to clarify the meaning of any symbol if needed. The axial direction is parallel to the rotor shaft and is in the same direction of the wind velocity.

The Reynolds number (Re) is first calculated so that we know which curve to use for the NACA airfoil performance curves, because the choice of the curve depends on that number. So, it is calculated as

$$\text{Re} = \frac{V\,c}{\nu} \quad (1)$$

The Reynolds number is a dimensionless value from which one can estimate the state of the flow, being laminar or turbulent [16]. It quantifies the ratio of the convective inertial force of a moving fluid to its viscosity-induced resistance for mixing [17]. After knowing the Re, we should be able to find the stall $C_L$ for the airfoil by using the suitable figures in the NACA family of airfoils [18]. Then, the actual design $C_L$ will be 85% of that value to have a 15% safety buffer away from stall. Because the air density is also known and the chord length is known, we should be able to calculate the lift force per unit span after knowing the value of the relative velocity at the particular location along the blade span.

An example of the $C_L$ curves for the NACA 0012 airfoil is shown in Figure 9 [18].

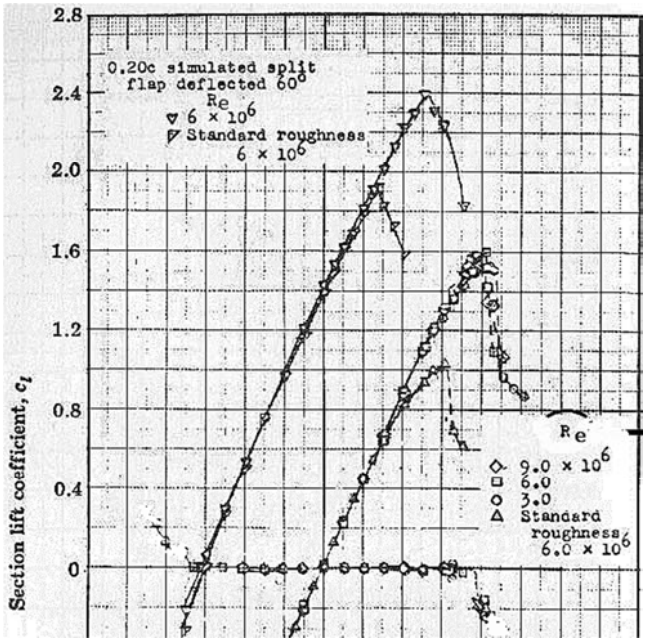

*Figure 9. Lift coefficient curves for NACA 0012 airfoil.*

Then, the angular velocity is obtained from the rpm as

$$\omega = \frac{2\pi \, rpm}{60} \quad (2)$$

Now, we select a radial distance along the blade span. For this particular location, we perform a sequence of calculations as presented in the following text to finally find the power per unit span at this radial location. This sequence should be repeated at a number of radial locations, and then a numerical integration is performed to calculate the power output from the entire blade span. This then is multiplied by the number of blades in the rotor to give the total power output from the rotor.

So, at a given radial distance ($r$) from the rotor center, we calculate

$$U = \omega \, r \quad (3)$$

$$\beta = \tan^{-1}\left(\frac{U}{V}\right) \quad (4)$$

$$W = \sqrt{(U^2 + V^2)} \quad (5)$$

$$L = \frac{1}{2}\rho W^2 \, c \, C_L \quad (6)$$

$$D = \frac{1}{2}\rho W^2 \, c \, C_D \quad (7)$$

Although the effect of the drag is small because the drag is about 2% only of the lift, we account for its effect where it decreases the tangential force which is mainly built by the lift force.

$$F_\theta = L \cos \beta - D \sin \beta \quad (8)$$

Finally, the power from the rotor per unit span at the selected radial distance ($r$) is

$$P_r = F_\theta \, U \quad (9)$$

Approximating $P_r(r)$ as a piecewise-liner function, and simplifying the integration $\int_{r_h}^{r_t} P_r(r)dr$, we get:

$$P = N_b [0.5 \, (P_{r1}+0) \, (r_1 - r_h) + 0.5 \, (P_{r2}+P_{r1}) \, (r_2 - r_1) + 0.5 \, (P_{r3}+P_{r2}) \, (r_3 - r_2)\ldots] \quad (10)$$

where $N_b$ is the number of blades. In the above equation, the zero refers to the estimated power per unit span at the hub. For the tip radius of the blade ($r_t$), we can use extrapolation from the two previous radial locations to estimate the power per unit span at that location. For example, if $P_{rt}$ is $P_{r4}$ then

$$P_{r4} = P_{r3} + \frac{P_{r3}-P_{r2}}{r_3-r_2}(r_4 - r_3) \quad (11)$$

## 5. Results

In this section, we present the calculation results of the wind turbine rotor we propose. We first present the input parameters or limits we considered as fixed inputs or recommended design data based on the review of typical wind turbines from sample manufacturers of a wind turbines in the same family we are interested in, as well as the peak mean annual wind speed recorded for Oman (in Thumrait). Beside these, we select values of other design variable from a reasonable range and then apply the equations presented in the previous section and check if the rotor output power is satisfactory. If the power is too low or too high, one or more of the design variables is changed until we reach a good final design. We have used the software package MATLAB by MathWorks [19] to perform the calculations quickly. At the end, we decided a rotor design and it is presented later. The MATLAB code is given in the Appendix.

### 5.1. Airfoil Section

We use the airfoil section NACA 0012 [18] for the rotor blade. It is a symmetric airfoil, making it easy to manufacture and thus cheaper than asymmetric (cambered) designs. This airfoil has been examined in other works as a candidate section for wind turbine blades [20-22]. The predicted performance for a small-scale wind turbine with this airfoil was comparable to that with the cambered airfoil NACA 4412, provided that it operates near the design rotational speed [23]. Instead of using a simplified curve for the lift coefficient as a function of the incidence [15], we used the actual curves published by the National Aeronautics and Space Administration (NASA) in the USA, which was previously called the National Advisory Committee for Aeronautics (NACA). There is more than one curve for this airfoil, depending on the Reynolds number of the flow. We found that our Reynolds number is below $10^6$, and thus will use the low-Re curve having circles in Figure 10. We then set the operational lift coefficient at 85% of the stall value to be safely far from the undesirable stall. Then, the operational incidence will be the incidence corresponding to that operational lift coefficient; both are marked in Figure 10.

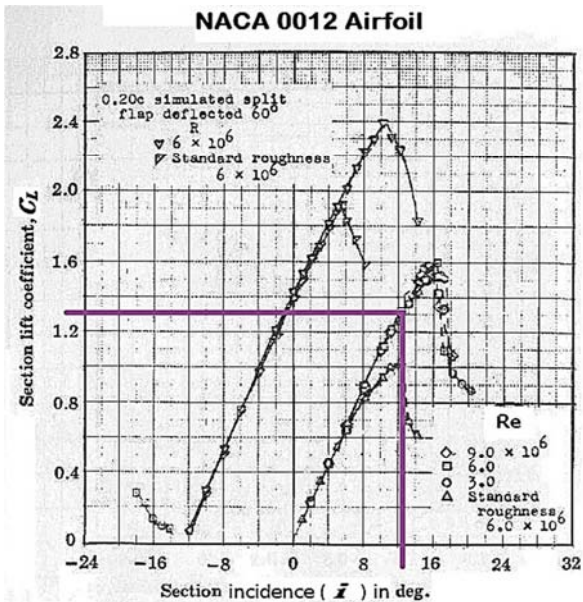

*Figure 10. Lift coefficient curves for NACA 0012 airfoil.*

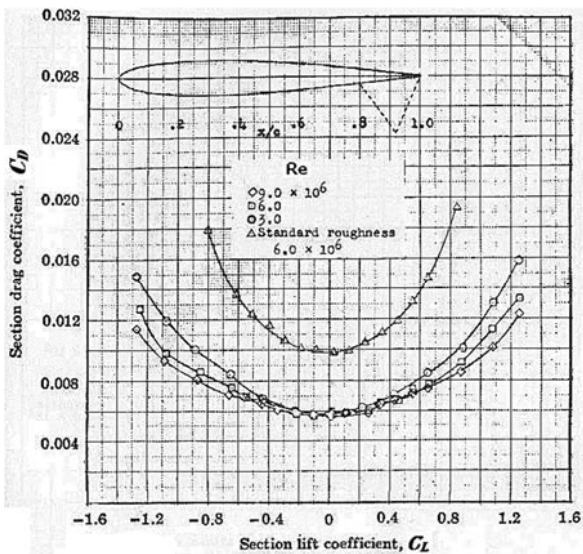

*Figure 11. Drag coefficient curves for NACA 0012 airfoil.*

There is another set of curves of the drag coefficient as a function of the lift coefficient. We use Figure 11 to estimate the drag coefficient for our NACA 0012 airfoil. Unlike the lift coefficient, where the curves at different Re were very close to each other at the operational point, the curves for the drag coefficient are quite separated and thus the effect of the Re is more important. We account for this by taking a higher value that lies above the low-Re curve because our operational Re is below the value of the curve, and thus the drag coefficient should be higher because it increases as the Re decreases.

### 5.2. Fixed Parameters

The fixed design parameters are:
1. Wind density: $1.22 kg/m^3$ [24].
2. Wind kinematic viscosity: $1.5 \times 10^{-5} m^2/s$ [25].
3. Wind speed: 6m/s.
4. Lift coefficient: 1.3.
5. Drag coefficient: 0.018.
6. Incidence: 12.5°.
7. Number of blades: 3.
8. Hub-to-tip diameter ratio: 5%.
9. Hub height: 80m.
10. Gearbox: two stages.
11. Blade material: epoxy.
12. Tower material: steel.

### 5.3. Design Variables

The design variables (and the recommended ranges) are:
1. Rotor diameter: 80 – 110m.
2. Rotational speed: 6 – 18rpm.
3. Chord length: up to 4m.

### 5.4. Final Design

We selected the following values of the design variables:
1. Rotor diameter: 80m.
2. Rotational speed: 15rpm.
3. Chord length: 3.5m.
Therefore, we have:
1. Hub radius: 2m.
2. Tip radius: 40m.
3. Span: 38m.
4. Re: $1.6 \times 10^6$.
5. Power output: 2.37MW.

The diameter is the smallest in the recommended range, which is a choice guided by a desire to avoid an increase in the structural stresses with the increased inertia, and also the increased noise with the increases tip speed.

When calculating the rotor power, 16 radial locations were placed along the blade with equal spacing between them. We found that this number is very sufficient and when we doubled it to 32, we obtained the same power result. If the efficiency of the gearbox and electric generator and any other mechanical or electrical components together is 85% [26, 27], then the electric power will be approximately 2MW, which is our target.

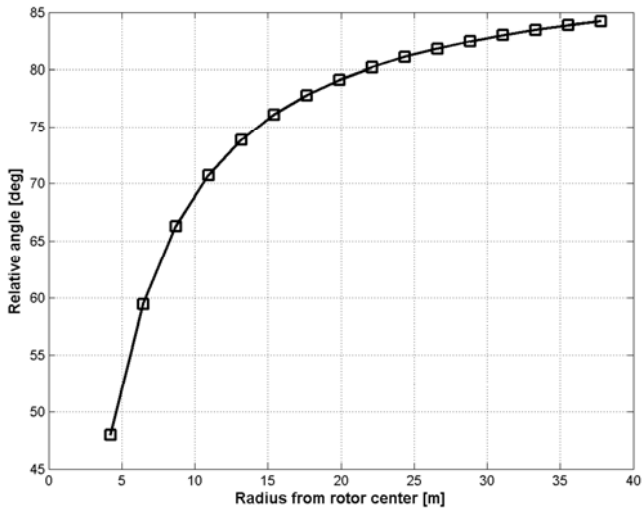

*Figure 12. Distribution of the relative angle along the blade.*

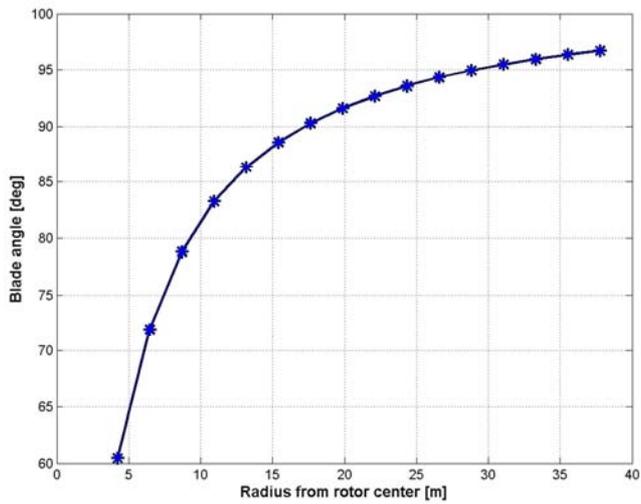

*Figure 13. Distribution of the blade angle along the blade.*

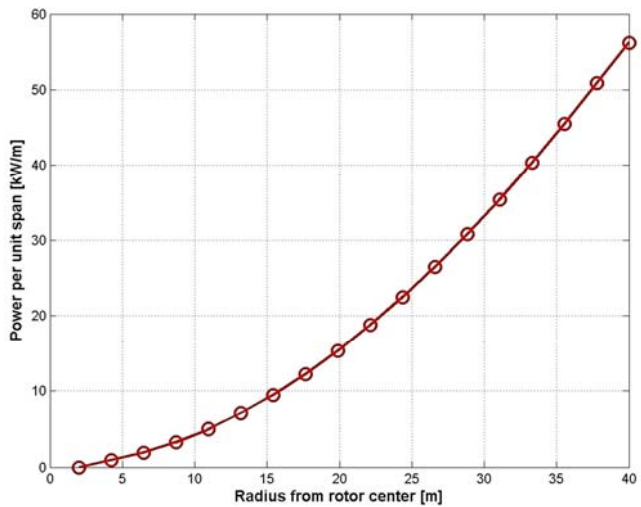

*Figure 14. Distribution of the power per unit span along the blade.*

Figure 12 shows the distribution of the relative flow angle along the blade. This angle increases to near 90° near the tip because of the very high tangential velocity there. In order to keep the incidence fixed, the blade must be twisted and the blade angle must also increase as we go to the tip, as shown in Figure 13.

The distribution of the power per unit span is given in Figure 14. The part of the blade near the tip is more important than the part near the hub because it has a much larger share in the rotor power.

Table 2 gives a summary of the calculated distributions along the blade radius for the blade angle and the power per unit span.

*Table 2. Radial distribution of the blade angle and the power per unit span.*

| Radial station number | Radius (m) | Blade angle (degrees) | Power per unit span (kW/m) |
|---|---|---|---|
| hub | 2 | - | 0 |
| 1 | 4.2 | 60.5 | 0.98 |
| 2 | 6.5 | 71.9 | 1.95 |
| 3 | 8.7 | 78.8 | 3.29 |
| 4 | 10.9 | 83.3 | 5.00 |
| 5 | 13.2 | 86.3 | 7.07 |
| 6 | 15.4 | 88.6 | 9.49 |
| 7 | 17.6 | 90.3 | 12.25 |
| 8 | 19.9 | 91.6 | 15.35 |
| 9 | 22.1 | 92.7 | 18.76 |
| 10 | 24.4 | 93.6 | 22.49 |
| 11 | 26.6 | 94.3 | 26.52 |
| 12 | 28.8 | 95.0 | 30.84 |
| 13 | 31.1 | 95.5 | 35.44 |
| 14 | 33.3 | 96.0 | 40.31 |
| 15 | 35.5 | 96.4 | 45.45 |
| 16 | 37.8 | 96.7 | 50.84 |
| tip | 40 | - | 56.22 |

## 6. Conclusions

Using a combination of a literature data, simple mathematical modeling, and numerical calculation, we proposed a coarse design of a horizontal-axis wind turbine, especially its rotor, so that it can produce 2MW of electricity when used in Thumrait, Dhofar, Sultanate of Oman or other place with a mean wind speed of 6m/s. The rotor radius is 40 m, the hub radius is 2m, and the hub height is 80m. The incidence (angle of attack) is 12.5 degrees, and a symmetric NACA 0012 airfoil is used in the blade section. The rotor has 3 blades and its diameter is 70m. We found that the blade (twist) angle should vary from about 60 degrees at the hub to about 97 degrees at the tip. The power-per-unit span increases nonlinearly from zero at the hub to about 56kW/m at the tip. One can improve this work by considering other details and components of the wind turbine, like the generator selection, gearbox design, the load on the tower, the control of the blade pitch. However, these points are not within the scope of this work.

## Nomenclature

*Abbreviations (in Alphabetical Order)*

BTU British thermal unit (1055J)

GCC Gulf Cooperation Council (union of the 6 Arab states of the Gulf)

Quad   quadrillion BTU ($10^{15}$ BTU)
TWh   terawatt-hour ($10^{12}$ Wh)

*Latin Symbols and Abbreviations (in Alphabetical Order)*

| | |
|---|---|
| b | span |
| C | chord |
| $C_L$ | lift coefficient |
| $C_D$ | drag coefficient |
| D | drag force |
| $F_\theta$ | tangential force |
| *i* | incidence |
| L | lift force |
| $N_b$ | number of blades |
| P | power from the rotor |
| $P_r$ | power per unit span |
| r | radius |
| $r_h$ | hub radius |
| $r_t$ | tip radius |
| Re | Reynolds number |
| rpm | revolutions per minute |
| U | tangential (rotational) velocity |
| V | absolute velocity |
| W | relative velocity |

*Greek Symbols (in Alphabetical Order)*

| | |
|---|---|
| β | relative angle (between relative velocity and the axial) |
| γ | blade angle (also called twist angle) |
| ν | kinematic viscosity (typically $1.5 \times 10^{-5}$ m$^2$/s for air) |
| ρ | density |
| ω | angular velocity |

## Appendix (MATLAB Computer Code for the Design)

```
V = 6;
rho = 1.22;
c = 3.5;
rt = 80/2;
rh = 0.05 * rt;
b = rt - rh;
rpm = 15;
Nb = 3;
CL = 1.3;
CD = 0.018;
Nr = 16;
omega = 2 * pi * rpm / 60;
list_r = zeros(1, Nr);
list_Pr = zeros(1, Nr);
for n = 1 : Nr
  r = n / (Nr + 1) * b + rh;
  list_r(n) = r;
  U = omega * r;
  W = sqrt(U^2 + V^2);
  beta = atand(U/V);
  L = 0.5 * rho * W^2 * c * CL;
  D = 0.5 * rho * W^2 * c * CD;
  F_theta = L * cosd(beta) – D * sind(beta);
  list_Pr(n) = F_theta * U;
end
P = Nb * 0.5 * (list_Pr(1) + 0) * (list_r(1) - rh);
P_rt=list_Pr(Nr) + (list_Pr(Nr)-list_Pr(Nr-1))/(list_r(Nr)-list_r(Nr-1))*(rt-list_r(Nr));
P = P + Nb * 0.5 * (P_rt + list_Pr(Nr)) * (rt - list_r(Nr));
for n = 1 : Nr-1
  P = P + Nb*.5*(list_Pr(n+1)+list_Pr(n))*(list_r(n+1)-list_r(n));
end
```